\documentclass[twocolumn]{aastex7}
\usepackage{amsmath}
\usepackage{booktabs}
\usepackage{fontawesome}
\usepackage{bm}
\usepackage{amsbsy}
\usepackage{fontawesome}
\usepackage{threeparttable}

\usepackage{soul, CJK}
\usepackage[utf8]{inputenc}

\usepackage{xspace}
\DeclareFixedFont{\ttb}{T1}{txtt}{bx}{n}{9} 
\DeclareFixedFont{\ttm}{T1}{txtt}{m}{n}{9}  
\usepackage{color}
\definecolor{deepblue}{rgb}{0,0,0.5}
\definecolor{deepred}{rgb}{0.6,0,0}
\definecolor{deepgreen}{rgb}{0,0.5,0}
\usepackage{listings}
\usepackage{gensymb}

\newcommand\pythonstyle{\lstset{
language=Python,
basicstyle=\ttm,
otherkeywords={self},             
keywordstyle=\ttb\color{deepblue},
emph={MyClass,__init__},          
emphstyle=\ttb\color{deepred},    
stringstyle=\color{deepgreen},
frame=tb,                         
showstringspaces=false            %
}}
\lstnewenvironment{pyt}[1][]
{
\pythonstyle
\lstset{#1}
}
{}

\newcommand{\R}{\mathbb{R}}
\newcommand{\N}{\mathbb{N}}

\newcommand{\g}{\,|\,}

\newcommand{\thetae}{\theta_{\rm E}}
\newcommand{\pie}{\pi_{\rm E}}
\newcommand{\ks}{K_{\rm s}}
\newcommand{\kss}{K_{\rm s, S}}
\newcommand{\hs}{H_{\rm S}}
\newcommand{\ksl}{K_{\rm s, L}}

\newcommand{\sa}{s_{\rm A}}
\newcommand{\ssb}{s_{\rm B}}

\usepackage{listings}

\graphicspath{{./}{figures/}}

\shorttitle{KMT-2018-BLG-0029Lb and OGLE-2019-BLG-0960Lb}
\shortauthors{Zhang et al.}

\begin{document}
\begin{CJK*}{UTF8}{gkai}

\title{KMT-2018-BLG-0029Lb and OGLE-2019-BLG-0960Lb: Mass Measurements for Two Super-Earth Microlensing Planets}

\correspondingauthor{Keming Zhang}

\author[0000-0002-9870-5695]{Keming Zhang (张可名)}
\altaffiliation{Schmidt AI in Science Postdoctoral Fellow}
\affiliation{Department of Astronomy and Astrophysics, University of California, San Diego, CA 92093, USA}
\affiliation{Hal{\i}c{\i}o\u{g}lu Data Science Institute, University of California, San Diego, CA 92122, USA}
\email[show]{kemingz@berkeley.edu}

\author[0000-0002-5029-3257]{Sean K. Terry}
\affiliation{Department of Astronomy, University of Maryland, College Park, MD 20742, USA}
\affiliation{Code 667, NASA Goddard Space Flight Center, Greenbelt, MD, 20771, USA}
\email{skterry@umd.edu}

\author{Joshua S. Bloom}
\affiliation{Department of Astronomy, University of California, Berkeley, CA 94720-3411, USA}
\email{joshbloom@berkeley.edu}

\author{B. Scott Gaudi}
\affiliation{Department of Astronomy, The Ohio State University, Columbus, OH 43210, USA}
\email{gaudi.1@osu.edu}

\author[0000-0001-9611-0009]{Jessica R. Lu}
\affiliation{Department of Astronomy, University of California, Berkeley, CA 94720-3411, USA}
\email{jlu.astro@berkeley.edu}

\begin{abstract}
KMT-2018-BLG-0029Lb and OGLE-2019-BLG-0960Lb were the lowest mass-ratio microlensing planets at the time of discovery.
For both events, microlensing parallax measurements from the Spitzer Space Telescope implied lens systems that were more distant and massive than those inferred from the ground-based parallax.
Here, we report on the detection of excess flux aligned to the event locations using Keck Adaptive Optics imaging, which is consistent with the expected brightness of main-sequence hosts under the ground-based parallax, but inconsistent with that predicted by Spitzer.
Based on the excess flux, ground-based parallax, and angular Einstein radius, we determine KMT-2018-BLG-0029Lb to be a $4.2\pm0.5 M_\oplus$ planet orbiting a $0.70\pm0.07 M_\odot$ host at a projected separation of $3.1\pm0.3$ au, and OGLE-2019-BLG-0960Lb to be a $2.0\pm0.2 M_\oplus$ planet orbiting a $0.40\pm0.03 M_\odot$ host at a projected separation of $1.7\pm0.1$ au.
We report on additional light-curve models for KMT-2018-BLG-0029 under the generalized inner-outer (offset) degeneracy, which were not reported in the original analysis.
We point out inconsistencies in the inner/outer labeling of the degenerate models in the lens and source planes, and advocate for the lens-plane convention, which refers to the planet being closer or further to the host star compared to the image it perturbs. 
Lastly, we discuss the possibility of breaking this degeneracy via ground concurrent observations with the Roman Space Telescope.
\end{abstract}


\section{Introduction}
KMT-2018-BLG-0029 (KB180029; \citealt{gould_kmt-2018-blg-0029lb_2020}) and OGLE-2019-BLG-0960 (OB190960; \citealt{yee_ogle-2019-blg-0960_2021}) were two high magnification microlensing events that led to the discovery of planets with exceptionally low mass ratios. Their underlying microlensing light curves exhibited both finite-source and microlensing parallax effects, each of which yields an independent mass-distance relationship for the lens system, thereby enabling the lens masses and distances to be relatively well constrained.

In particular, the finite-source effect informs the angular Einstein radius of the lens system,
\begin{equation}
    \thetae=\sqrt{\kappa M_{\rm L} \pi_{\rm rel}},
\end{equation}
where $\pi_{\rm rel}=\pi_{\rm L}-\pi_{\rm S}$ is the lens-source relative parallax and
\begin{equation}
    \kappa=\dfrac{4G}{c^2{\rm au}}\simeq8.144 {\rm mas}/M_\odot.
\end{equation}
On the other hand, the microlensing parallax is defined as the lens-source relative parallax ($\pi_{\rm rel}$) in units of the angular Einstein radius:
\begin{equation}
    \pi_{\rm E}=\dfrac{\pi_{\rm rel}}{\theta_{\rm E}}=\sqrt{\dfrac{\pi_{\rm rel}}{\kappa M_{\rm L}}}.
\end{equation}
Therefore, the mass of the lens can be derived as $M_{\rm L}=\theta_{\rm E}/(\kappa\cdot\pi_{\rm E})$, while the lens parallax is $\pi_{\rm L}=\pi_{\rm rel}-\pi_{\rm S}=\pi_{\rm E}\cdot\theta_{\rm E}-\pi_{\rm S}$. For nearby disk lenses, the uncertainty in the source distance generally does not constitute a dominant error term, and is adopted as $D_{\rm s}\simeq8$ kpc and $\pi_{\rm s}\simeq0.125$ mas.

For both events, the microlensing parallax is measured via both the annual and satellite parallax effects, and the satellite parallax measured by Spitzer is substantially smaller than the annual parallax, thereby leading to discrepancies in the predicted lens properties. Spitzer microlensing parallax can be susceptible to systematic errors (e.g., \citealt{koshimoto_evidence_2020}), particularly when Spitzer does not detect the peak of the light curve, which is indeed the case for both events. For OB190960, the Spitzer parallax would have implied a main-sequence-lens apparent brightness that is significantly brighter and bluer than the seeing-limited blend flux, which sets an upper limit to the lens flux. For this reason, \cite{yee_ogle-2019-blg-0960_2021} did not include the Spitzer parallax in their calculation of the physical parameters of the lens.

As for KB180029, \cite{gould_kmt-2018-blg-0029lb_2020} noted the presence of correlated residuals in at least parts of the Spitzer data. They carefully inspected the full Spitzer data and only included a subset for their analysis, leading to a Spitzer-only parallax of $\pie\simeq0.116\pm0.007$. This value was reported to be consistent with the annual parallax under the $u_0>0$ model ($\pie=0.151\pm0.080$), but mildly inconsistent under the $u_0<0$ model ($\pie=0.280\pm0.126$). Here, $u_0$ is the impact parameter, and its sign indicates the direction of the source's trajectory under the ecliptic degeneracy \citep{smith_acceleration_2003,jiang_ogle-2003-blg-238_2004}.

Furthermore, the expected main-sequence lens brightness resulting from the joint parallax constraint from the ground and Spitzer was reported to be brighter than the seeing limited blend flux at the 1.5-$\sigma$ level for the $u_0<0$ model. Because of this, the authors considered the $u_0<0$ model disfavored.
On the other hand, the reported consistency under the $u_0>0$ model is contingent on the assumption that the blended flux primarily reflects the lens flux, and that the lens lies behind the full column of dust ($A_I=3.39$) inferred for the source star.

Of course, the alternative explanation is that either the curated Spitzer data remain affected by systematics, or perhaps the lens star is a dark stellar remnant. Both possibilities may be further investigated by resolving the event location with ground-based adaptive optics (AO) imaging in the infrared, setting a tighter constraint on the lens brightness. The stellar remnant hypothesis could be confirmed by AO observations if the excess flux aligned to the source location is fainter than the faintest main-sequence brightness allowed by the light-curve models.
The infrared pass band offers an additional advantage of having much lower extinction, greatly reducing the dependency on the assumed lens extinction, especially for KB180029.

As part of our 2023A Keck program (U152; PI: Joshua Bloom, Science-PI: Keming Zhang), we have observed multiple planetary microlensing events with measured lens masses and distances, in order to search for white-dwarf planet hosts. Our program has led to the discovery of the first terrestrial planet orbiting a white dwarf \citep{zhang_earth-mass-1_2024}, and this paper reports on two planets for which we detect the expected main-sequence lens flux, namely KB180029 and OB190960.

This work is organized as follows. In Section 2, we report on additional light-curve models for KB180029 under the generalized inner-outer (offset) degeneracy \citep{zhang_ubiquitous_2022,zhang_mathematical_2022}, which were not reported in the original analysis. In Section 3, we report on our observations and derive the excess flux.
In Section 4, we determine the physical parameters for both systems based on all available constraints. We discuss our results in Section 5.

\section{Light-Curve Models}

We first report on additional light-curve models for KB180029 that were not reported in the original analysis of \cite{gould_kmt-2018-blg-0029lb_2020}. The published models for KB180029 are firmly in the resonant caustic regime with projected separation $s=1.000\pm0.002$, in units of the angular Einstein radius. In this regime, both the ``close-wide'' degeneracy for central caustics \citep{griest_use_1998} and the ``inner-outer'' degeneracy for planetary caustics \citep{gaudi_planet_1997} are known to break down.
The possibility that these degeneracies may merge in the resonant regime was first suggested by \cite{yee_ogle-2019-blg-0960_2021} in their very analysis of OB190960. Subsequently, \cite{zhang_ubiquitous_2022} showed that the central and planetary caustic degeneracies are in fact limiting cases of a universal magnification degeneracy, which we refer to as the generalized inner-outer degeneracy in this work.

This universal magnification degeneracy was originally introduced as the offset degeneracy by \cite{zhang_ubiquitous_2022}, where the term offset referred to deviations from the $s\leftrightarrow1/s$ invariance of the central caustic degeneracy. Nevertheless, \cite{zhang_perturbative_2023} later found the offset degeneracy to be more fundamentally connected to the planetary caustic degeneracy.
The main result of \cite{zhang_perturbative_2023} is that regardless of the caustic structure associated with the light-curve anomaly, the planet only perturbs one of the major/minor images at a time, and the action of the planet is accurately described by a Chang-Refsdal Lens, which captures the behavior of isolated planetary caustics.
The implication is that essentially all planetary light-curve anomalies could be characterized by ideal, isolated planetary caustics (i.e., Chang-Refsdal caustics), and thus the planetary caustic degeneracy naturally generalizes as a universal magnification degeneracy.

For any caustic topology, the generalized inner-outer degeneracy expects a pair of projected separation solutions ($s_{\rm A,B}$) satisfying
\begin{equation}
    \dfrac{(\sa-1/\sa)+(\ssb-1/\ssb)}{2}=\dfrac{u_0}{\sin(\alpha)},
\end{equation}
where $u_0/\sin(\alpha)\simeq0.025$ is the intercept of the source trajectory on the star-planet axis (Figure 1), and $s_{\rm A,B}-1/s_{\rm A,B}$ is the location of the planetary caustic. In the resonant caustic regime, two off-axis planetary caustic cusps can still be identified at this location, marked by the vertical dashed lines in Figure 1.
Therefore, we derive the degenerate solution to be $s=1.027\pm0.002$. As illustrated in Figure 1, the source trajectory passes exactly half way between the planetary caustic cusps for the degenerate models, as required by Equation 4.

\begin{figure}[t]
    \centering
    \includegraphics[width=\columnwidth]{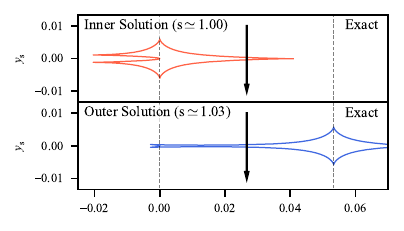}
    \includegraphics[width=\columnwidth]{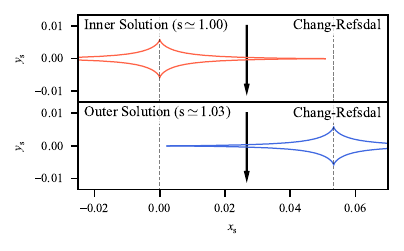}
    \caption{Caustic representations for the inner and outer solutions for KMT-2018-BLG-0029. The vertical dashed lines mark the locations of the planetary caustic/cusp located at $s-1/s$, which are equidistant to the source trajectory marked with the downwards arrow. Top: exact caustic representation. Bottom: caustic representations under the Chang-Refsdal Lens approximation, where the inner and outer solutions are shown to be exactly symmetrical.}
    \label{fig:enter-label}
\end{figure}

\begin{figure}[t]
    \centering
    \includegraphics[width=\linewidth]{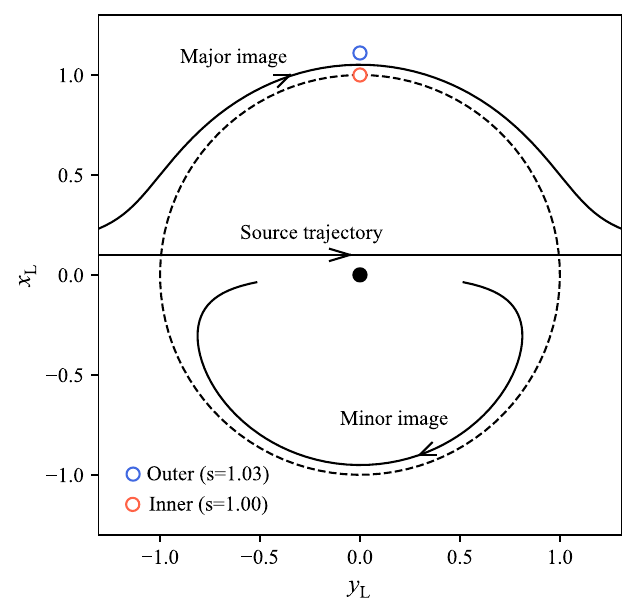}
    \caption{Illustration of KMT-2018-BLG-0029 in the lens plane, where the planet is shown to perturb the major image from two possible locations, marked by the red and blue open circles. The Einstein radius is shown as a dashed circle. The trajectories of the source star, major image, and minor image are shown below their respective labels, with the direction of the trajectories indicated by the arrows. Note that the axes are rotated 90 degrees with respect to Figure 1.}
    \label{fig:enter-label}
\end{figure}

The generalized inner-outer degeneracy is exact when the source trajectory crosses perpendicularly to the star-planet axis, which is indeed the case for KB180029. As illustrated in the lower panels of Figure 1, vertical source trajectories result in exact symmetry under the Chang-Refsdal Lens representation, with the Chang-Refsdal caustic located at $s-1/s$ and the shear evaluated at the same unperturbed image locations for both models. We verified that the light curves under the degenerate models differ by less than 0.05\%. Therefore, the remaining set of microlensing parameters are expected to be identical for the degenerate models. We assign equal weights to the inner/outer projected separations differing by 2.7\% in deriving the physical lens properties in Section 4.

\subsection{Labeling: Inner vs. Outer}

Both the original and generalized inner-outer degeneracies describe an ambiguity as to ``whether the planet lies closer to or farther from the star than does the position of the image that it is perturbing[,]'' as originally conceived by \cite{gaudi_planet_1997}.
As illustrated in Figure 2, the planet perturbs the major image, and we label the degenerate solutions inner ($s\simeq1.00$) and outer ($s\simeq1.03$) based on the planet being closer or further to the host star as compared to the major image.
Thus, the the inner/outer labels we adopt refer to the lens plane.

In comparison, the inner/outer labels have also been applied in the source plane, which refer to the source trajectory being interior or exterior to the planetary caustic, with respect to the central caustic.
The terms inner/outer are perhaps first used by \cite{bennett_planetary_2012} to refer to the inner/outer cusps of the planetary caustic.
Later, both the lens- and source-plane labeling were interchangeably used by \cite{han_moa-2016-blg-319lb_2018} for the minor image perturbation event MOA-2016-BLG-319, which states ``[because] the source trajectory passed the inner region of the planetary caustic with respect to the planet host, we refer to this solution as `inner solution.' The same paper later states ``the planet is located inside the minor image (with respect to the host) for the `inner solution'.''
In other words, the lens-plane inner/outer labels refer to the planet's location, whereas the source-plane labels refer to the source star's location.

\begin{figure*}[t]
    \centering
    \includegraphics[width=\textwidth]{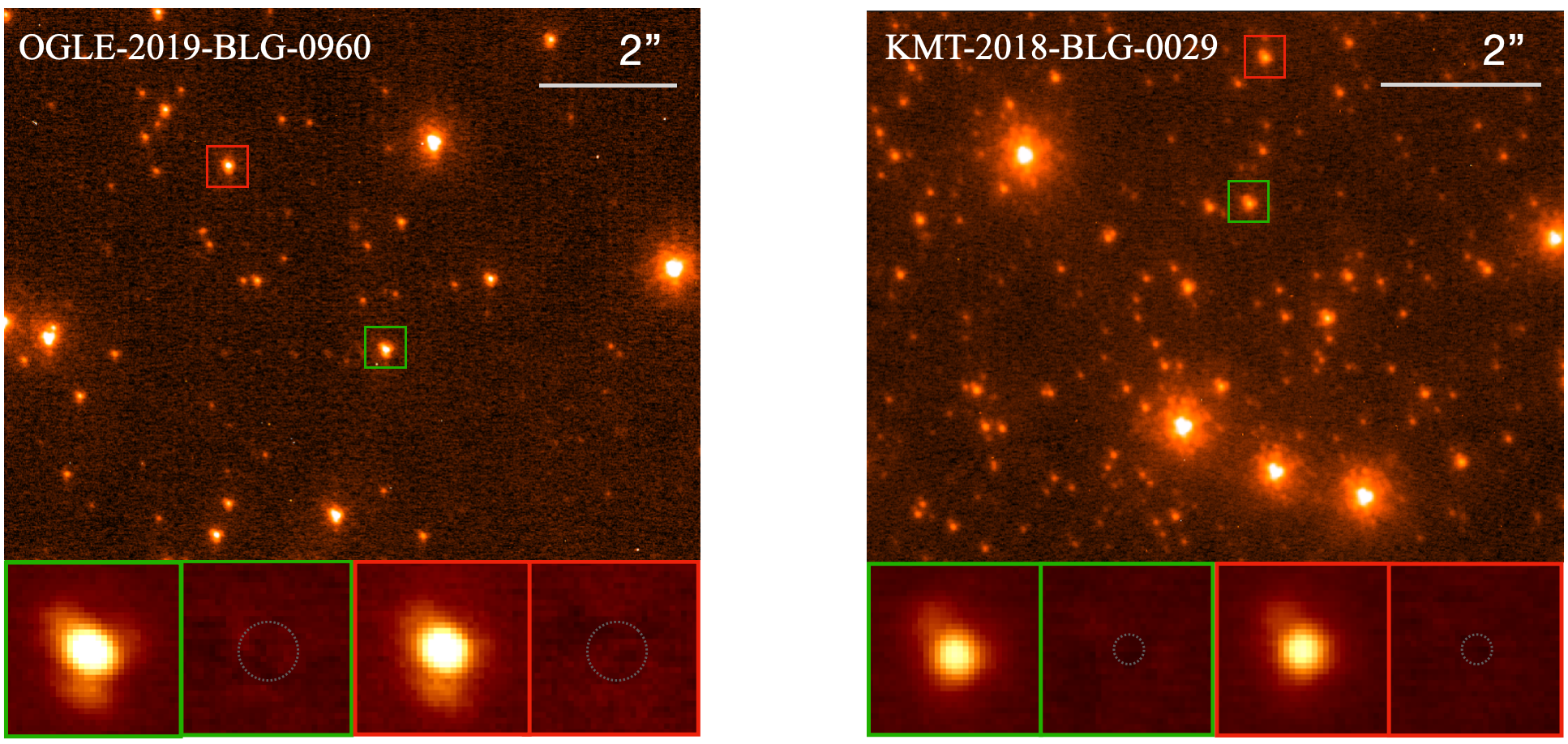}
    \caption{Keck NIRC2-LGS images of the two events. North is up and East is to the left. The green and red boxes indicate the target and comparison stars, respectively. Insets show zoomed-in views of each star, including the original image and the residuals following PSF subtraction, which are shown on the identical flux scale to facilitate comparison. The gray dotted circles in the residual images show the predicted lens-source separation.}
    \label{fig:enter-label}
\end{figure*}

However, while the lens/source-plane conventions are consistent for minor image perturbations, we find that they are contradictory for major image perturbations.
By comparing Figures 1 \& 2, we may see that the outer planet location ($s\simeq1.03$) is technically the inner solution in the source plane, as the source trajectory crosses in between the would-be\footnote{The resonant caustic would split into central and planetary caustics for a slightly smaller mass ratio at $s\simeq1.03$.} central and planetary caustics.
On the other hand, the inner planet location would have had the source trajectory passing the outer region of the wide planetary caustic.
However, in our specific case, there is actually no conceptual basis to refer to the $s\simeq1.00$ solution as the outer solution in the source plane, as both the would-be central and planetary caustics are co-located at the coordinate origin.

As the inner/outer labels in the source plane rely upon the idea of caustic resemblance that holds only for isolated planetary caustics, we advocate for the lens plane convention instead.
In the lens plane, the terms inner and outer always refer to the planet being closer or further to the host star, making it consistent with similar usages in the exoplanet literature beyond microlensing (e.g., inner-outer connection; \citealt{zhu_exoplanet_2021}).
In comparison, the source-plane convention would refer to an inner planet location as the outer solution (under major image perturbations), which becomes rather counter-intuitive.
The lens-plane convention also makes explicit the origin of the inner-outer degeneracy in symmetries under the perturbative picture, with the planet being inside or outside the image being perturbed.

\section{Lens Flux Constraints}
We observed the locations of KB180029 and OB190960 using the NIRC2 camera on the Keck-II telescope on May 25, 2023 (UT) under program U152 (PI: Joshua Bloom; Science-PI: Keming Zhang). For each target, we acquired one set of shallow images for absolute photometric calibration, and one set of deep images for relative photometry on the target (Figure 3).
For KB180029, six deep images were acquired with the narrow camera (0.01''/pixel), each with 30 seconds of exposure. Five calibration images were taken with the wide camera (0.04''/pixel) where each image consists of 50 co-adds of 0.181 seconds of exposure.

As for OB190960, 18 deep images are taken with the wide camera, each with 30 seconds of exposure. Ten calibration images were taken with the wide camera, each with 20 coadds of 0.5-second integrations. Due to a malfunction of the AO system, all images are centered on the tip-tilt (TT) natural guide star, which complicated the use of the narrow camera for OB190960 for which the TT star was far from the target. Nevertheless, we were able to acquire three images of OB190960 using the narrow camera with 30 seconds of exposure each, by manually offseting from the TT star.
The images are non-linearity corrected \citep{metchev_palomarkeck_2009}, sky-subtracted, flat-fielded, and then stacked.

We identify the KB180029 location using the magnified source location along with a list\footnote{\url{http://kmtnet.kasi.re.kr/ulens/data/KMT-2018-BLG-0029.CMD}} of reference stars published by \cite{gould_kmt-2018-blg-0029lb_2020}. The target is located at (625.57, 781.94) in the narrow image and (543.62, 577.60) in the wide image, as marked by the green box in Figure 3. The location of OB190960 was unambiguously identified due to the absence of confounding nearby sources.

We perform aperture photometry with a radius of 11 pixels (0.44'') on the wide image and a radius of 18 pixels (0.18'') on the narrow image using the \textit{photutils} package \citep{bradley_astropyphotutils_2022}. We calibrate the Keck instrumental magnitude to the VVV system by cross-matching
to the VVV 1''-radius aperture photometry (DR4; \citealt{minniti_vista_2010}) and applying inverse-variance weighted linear regression.
The variance includes both the photometric uncertainties from Keck/VVV, and the intrinsic zero point uncertainty, the latter of which is estimated iteratively from the residual sum of squares.
The calibrated magnitudes are listed in Table 1.

From the published lens-source proper motion for each event, we estimate the lens-source separation to be approximately 25 mas for KB180029 and 48 mas for OB190960. Given the image quality, the PSF FWHM of 70 mas, and the predicted lens-source flux ratios (Table 1), it would be difficult to spatially resolve the lens and source stars.
With this in mind, we further analyzed the narrow-camera images using the DAOPHOT-MCMC routine \citep{stetson_daophot_1987,terry_moa-2009-blg-319lb_2021} to perform PSF modeling and subtraction.
For each event, we constructed an empirical PSF model based on bright stars located near the target. This model was then applied to both the target star and a comparison star, both of which were excluded from the set of PSF stars.

As shown in Figure 3, the PSFs display a three-lobed trefoil aberration due to imperfect AO correction, which are consistent between the target and comparison stars for the same image. The PSF subtractions resulted in mild spatially correlated residuals with similar patterns and amplitudes for the target and comparison stars, which is primarily attributed to limitations in the PSF model due to spatial variations of the PSF across the image. For example, both residuals for KB180029 suggest slight over-subtraction at the PSF core, which indicates that the PSF model is more concentrated at the core. While OB190960 displays weak residuals along the North-East and South directions, they overlap with the trefoil aberration, whereas over-subtractions at similar amplitudes are also seen for the reference star towards the North-East direction.
Therefore, we conclude that there is insufficient evidence for an elongated PSF at the event locations.
We assume that any observed flux in excess of the derived source flux is contributed by the primary lens star, and explore alternative possibilities in Section 5.

\begin{figure*}[t]
    \centering
    \includegraphics[width=0.49\textwidth]{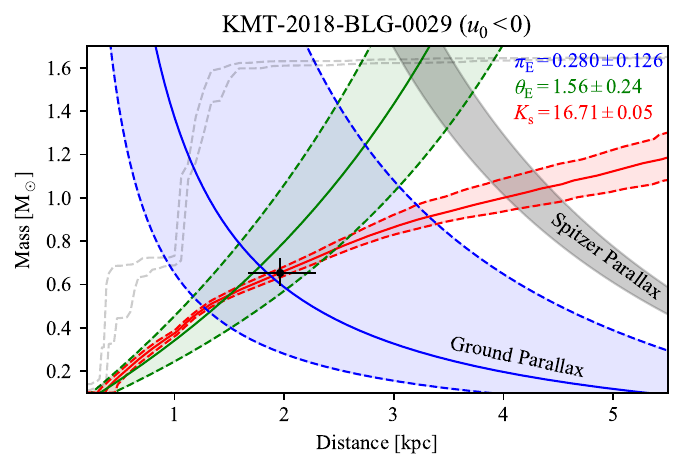}
    \includegraphics[width=0.49\textwidth]{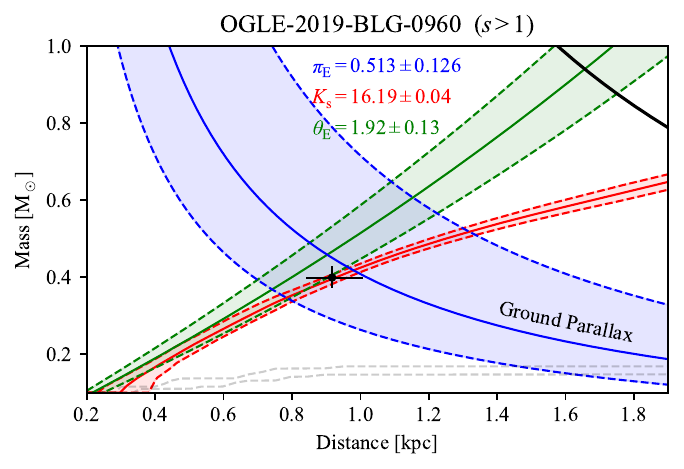}
    \includegraphics[width=0.49\textwidth]{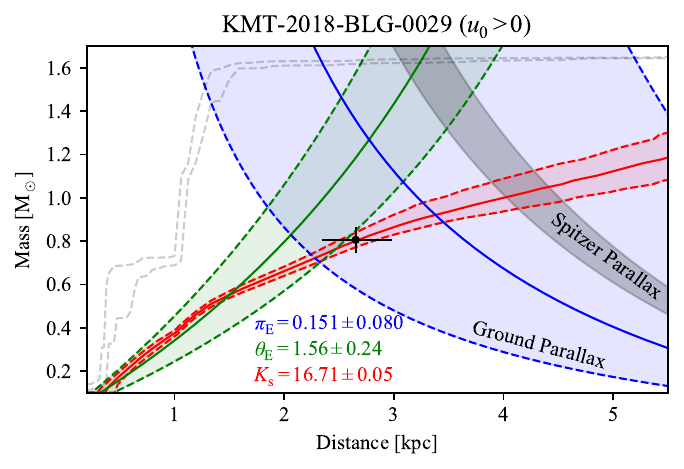}
    \includegraphics[width=0.49\textwidth]{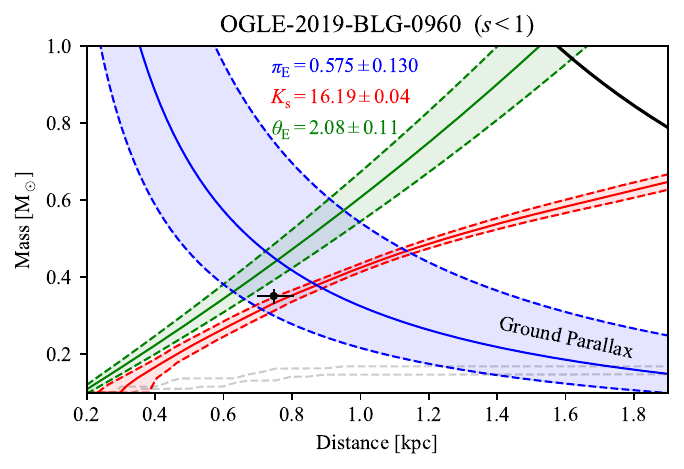}
    \caption{
    Constraints on the lens mass and distance for the two lens systems.
    \textbf{KMT-2018-BLG-0029L:} Results for the $u_0<0$ (top left) and $u_0>0$ (bottom left) microlensing parallax models.
    \textbf{OGLE-2019-BLG-0960L:} Results for the $s>1$ (top right) and $s<1$ models (bottom right), each combining $u_0>0$ and $u_0<0$ microlensing parallax solutions.
    In each panel, the shaded regions represent the 68\% confidence intervals of the mass-distance relationships for the lens system as derived from parallax constraints from ground ($\pie$, blue) and Spitzer (gray; only shown for KB180029), the angular Einstein radius ($\thetae$, green), and the observed $\ks$-band lens flux (red). For OB190960, the Spitzer parallax is shown as the black curve to the upper right, as no uncertainties were reported in the original analysis. The median posterior values for the lens mass and distance are shown as the black dot with the error bar indicating the nominal 68\% confidence intervals. The gray dashed lines (without shade) indicate the 68\% confidence interval of partial extinction as a function of distance, with the vertical span of each panel corresponding to 0\% (bottom edge) to 100\% (top edge) of the total extinction experienced by the source star.
    }
\end{figure*}

\begin{table}[t]
    \centering
    \begin{tabular}{c|cc}
         & KB180029&OB190960\\
         \hline
         $K_{\rm s,total}$&$16.39\pm0.03$&$16.00\pm0.03$\\
         $K_{\rm s,S}$&$17.87\pm0.09$&$17.98\pm0.07$\\
         $K_{\rm s,L}$&$16.71\pm0.05$&$16.19\pm0.04$\\
         \hline
    \end{tabular}
    \caption{Keck observations and derived source and lens brightnesses in the $\ks$ band.}
    \label{tab:my_label}
\end{table}

We then derive the source brightnesses in the $\ks$ band. For KB180029, the source brightness was directly measured in the $H$-band as $\hs=18.24\pm0.08$. We derive its $\ks$-band apparent magnitude using its intrinsic $(H-\ks)_0$ color and reddening $E(H-\ks)$. From Table 3 of \cite{gould_kmt-2018-blg-0029lb_2020}, we derive $(H-\ks)_{0,S}=0.06\pm0.03$. We derive the extinction from the reddening map of \cite{gonzalez_reddening_2012}, which gives $E(J-\ks)=0.95\pm0.12$ along this line-of-sight.
Assuming the extinction law of \cite{nishiyama_interstellar_2009}, we derive $E(H-\ks)=0.327\times E(J-\ks)=0.31\pm0.04$.
We therefore derive $\kss=17.87\pm0.09$ and thus $\ksl=16.71\pm0.05$, as listed in Table 1.

For OB190960, \cite{yee_ogle-2019-blg-0960_2021} reported $(I-H,I)_{\rm 0,S}=(0.92,18.85)\pm(0.03,0.05)$, which we convert to $(I-K)_{\rm 0,S}=0.99\pm0.03$ using \cite{bessell_jhklm_1988}. From \cite{gonzalez_reddening_2012} and \cite{nishiyama_interstellar_2009}, we derive $A_{\ks}=0.528\times E(J-\ks)=0.12\pm0.04$. We then derive $\kss=17.98\pm0.07$ and $\ksl=16.19\pm0.04$.

\section{Lens Properties}

We convert the derived $\ks$ lens brightness into a mass-distance relationship using the MESA \citep{paxton_modules_2011} Isochrones and Stellar Tracks (MIST; \citealt{choi_mesa_2016,dotter_mesa_2016}). The lens's apparent brightness depends on the mass, distance, age, metallicity, and extinction. For the same stellar mass, higher metallicity and younger age generally correspond to lower luminosity for main-sequence stars. Since the age and metallicity of the lens star is unconstrained, we consider a grid of isochrones with initial metallicity of $[\mathrm{Fe}/\mathrm{H}] \in [-0.25, 0.35]$ and age between 1 Gyr to 10 Gyr. The range of metallicity is adopted from the observed metallicity distribution at galactocentric radius $5<R<7$ kpc and height $|z|<0.5$ kpc \citep{hayden_chemical_2015}.

We derive the lens extinction as a fraction of the total $\ks$-band extinction inferred for the source star using the DECaPS 3D dust map \citep{zucker_deep_2025} for KB180029 and Bayestar19 3D dust map \citep{green_3d_2019} for OB190960, via the \texttt{dustmaps} python package \citep{m_green_dustmaps_2018}. Both dust maps are probabilistic, allowing us to incorporate extinction uncertainties in deriving the lens physical properties. As it turned out, KB180029 is nearly behind the full dust column towards the source, whereas OB190960 is nearly in front of the full dust column (Figure 4).

To derive the full posterior distribution of the lens mass  and distance, we jointly sample the likelihoods of the lens flux, ground-based parallax, and angular Einstein radius constraints.
The lens flux likelihood is constructed by taking the mean and standard deviation of the absolute magnitude for the aforementioned set of stellar isochrones varying in age and metallicity, before adding the uncertainties in the extinction and observed flux in quadrature.

For KB180029, we derive the lens physical properties separately for the $u_0>0$ and $u_0<0$ models, which have distinct parallax constraints.
As illustrated in Figure 4, the $\ks$-band lens brightness is too faint to be consistent with that implied by the Spitzer parallax for both models.
Therefore, the Spitzer parallax is most likely affected by systematics beyond the extent identified by \cite{gould_kmt-2018-blg-0029lb_2020}.

As shown in Figure 1 of \cite{gould_kmt-2018-blg-0029lb_2020}, the adopted Spitzer data presumed to be free of systematics can be separated into two groups: 15 epochs taken 6--12 days after the peak of the event (from Earth), and 14 epochs taken two years after the event, which sets the baseline. Due to this setup, the inferred satellite parallax is highly sensitive to the slope of the light curve over the 6--12 day post-peak period, which is unreliable in the presence of correlated noise.

Nevertheless, since parts of the Spitzer data are taken close to the event peak from ground, even one single measurement can be quite informative of the microlensing parallax \citep{gould_cheap_2012}. As discussed in the previous work, a single Spitzer measurement 6 days after the peak would constrain the microlensing parallax to a range of
\begin{equation}
    \pie = \dfrac{1\,\text{au}}{D_\perp}(u_{\rm spitzer}\pm u_\oplus),
\end{equation}
where $u_{\rm spitzer}\sim0.203$ and $u_\oplus\sim0.044$ are the lens-source projected separation as seen from Spitzer and ground, and $D_\perp\sim1.3$ au is the projected Earth-Spitzer separation, all of which defined at the measurement epoch and directly quoted from \cite{gould_kmt-2018-blg-0029lb_2020}. Therefore, a single Spitzer epoch would allow for satellite parallax as large as $\pie\simeq0.19$, which becomes compatible with the $\ks$ lens flux.

Given complications with the Spitzer data and the over-constrained nature of the problem, we do not include the Spitzer parallax in deriving the lens properties. Since the lens flux only marginally favor the $u_0<0$ model by $\Delta\chi^2\simeq1$ and all but the parallax constraint are identical across the two models, we average the parallax measurements into
\begin{equation}
    \pie=0.20^{+0.14}_{-0.11},
\end{equation}
where the resulting lens physical properties are listed in Table 2. Note that the combination of the lens flux and angular Einstein radius essentially sets an upper limit to the lens mass and distance, whereas the large parallax of the $u_0<0$ model sets a lower limit to the lens distance (and thus mass).
By substituting the Spitzer parallax with the lens flux, we find that the lens system is approximately 40\% less massive and closer than originally reported by \cite{gould_kmt-2018-blg-0029lb_2020}.

As for OB190960, the $u_0>0$ and $u_0<0$ models have ground-based microlensing parallax constraints that are consistent with each other to within 1-$\sigma$. On the other hand, the $s<1$ models have substantially larger measured angular Einstein radius ($\thetae\simeq2.1\pm0.1$ mas) compared to the $s>1$ models ($\thetae\simeq1.9\pm0.1$ mas). Therefore, we derive the system properties separately for the $s<1$ and $s>1$ models, for which the $u_0>0$ and $u_0<0$ solutions are combined.

\begin{table}[t]
    \centering
    \begin{tabular}{c|c|c|c|c}
         & $M_{\rm L}$ [$M_\odot$]&$M_{\rm p}$ [$M_\oplus$] &$a_{\perp}$ [au]&$D_{\rm L}$ [kpc] \\
         \hline
         $u_0<0$&${0.65}_{-0.06}^{+0.07}$& ${3.92}_{-0.39}^{+0.46}$& ${2.84}_{-0.28}^{+0.29}$& ${1.96}_{-0.30}^{+0.33}$\\
         $u_0>0$&${0.81}_{-0.06}^{+0.06}$& ${4.82}_{-0.56}^{+0.52}$& ${3.44}_{-0.24}^{+0.26}$& ${2.65}_{-0.30}^{+0.33}$\\
         \hline
         \textbf{combined}&${0.70}_{-0.07}^{+0.08}$& ${4.17}_{-0.41}^{+0.56}$& ${3.06}_{-0.31}^{+0.29}$& ${2.19}_{-0.33}^{+0.35}$\\
    \end{tabular}
    \caption{Lens physical properties for KB180029 derived from the lens flux, angular Einstein radius, and ground-based parallax. Reported values are median values and nominal 68\% confidence intervals.}
    \label{tab:my_label}
\end{table}

\begin{table}[t]
    \centering
    \begin{tabular}{c|c|c|c|c}
        & $M_{\rm L}$ [$M_\odot$]&$M_{\rm p}$ [$M_\oplus$] &$a_{\perp}$ [au]&$D_{\rm L}$ [kpc] \\
        \hline
        ($s<1$) &${0.35}_{-0.02}^{+0.02}$& ${1.48}_{-0.10}^{+0.11}$& ${1.39}_{-0.08}^{+0.08}$& ${0.75}_{-0.05}^{+0.06}$\\
        $\mathbf{s>1}$ &${0.40}_{-0.03}^{+0.03}$& ${1.95}_{-0.16}^{+0.14}$& ${1.67}_{-0.12}^{+0.14}$& ${0.92}_{-0.08}^{+0.09}$\\
    \end{tabular}
    \caption{Lens physical properties for OB190960 derived from the lens flux, angular Einstein radius, and ground-based parallax. The $u_0>0$ and $u_0<0$ parallax solutions are combined for each of the $s>1$ and $s<1$ solutions. The parenthesis around $s<1$ indicates that this model is rejected. Reported values are median values and nominal 68\% confidence intervals.}
    \label{tab:my_label}
\end{table}

As shown in Figure 4, the larger angular Einstein radius for the $s<1$ model is visibly inconsistent with the lens flux. We quantify this inconsistency in terms of the expected and observed lens brightnesses. The combination of $\thetae$ and $\pie$ predicts a lens brightness of $K_L=15.2\pm0.3$, resulting in a 3-$\sigma$ tension with the measured value of $K_L=16.19\pm0.04$. On the other hand, the $s>1$ light-curve models predict $K_L=15.7\pm0.4$, which is consistent with the observed value at approximately 1-$\sigma$. Therefore, we reject the $s<1$ model in favor of the $s>1$ model, with the derived lens properties listed in Table 3.

\section{Discussion}

We have observed the locations of the microlensing events KMT-2018-BLG-0029 and OGLE-2019-BLG-0960 using Keck Adaptive Optics to investigate the mass and distance of the lens systems. For both events, we measure an excess flux that is consistent with the main-sequence lens brightness expected from the ground-based light curve models. Based on the measured excess flux, we report on the refined or revised properties of the lens system.

For KB180029, we have reported on two additional light-curve models under the generalized inner-outer (offset) degeneracy that were not reported in the original analysis, and point out inconsistencies in the inner/outer labeling of the degenerate models under major image perturbations. We advocate for the lens plane convention, with inner/outer indicating the position of the planet with respect to the microlensing image being perturbed. Under this convention, the inner solution is always the close separation solution, with the outer solution being the wide separation solution.

We now explore whether the Keck excess flux could reflect a binary companion to the lens or source instead, both of which would most likely require the primary lens itself to be a dark stellar remnant, namely a white dwarf. Therefore, these scenarios are \textit{a priori} disfavored given the relative scarcity of white dwarf lenses. Furthermore, the projected separations of the hypothetical lens/source companions are restricted to a narrow range given that a) they weren't close enough to produce a observable signature in the light curves and b) they weren't far enough to be resolved from source star in the Keck images.

For the lens-companion scenario, the presence of a wide binary companion with $s\gg1$ induces an additional central caustic of size $\simeq q/s^2$ \citep{dominik_binary_1999}, which spatially coincides with the resonant caustic induced by the planet. To avoid being detected, this binary-induced caustic must be well within the central region probed by the source trajectory. Thanks to the high magnification nature of both events, we may rule out lens companions with projected separations within approximately $s\lesssim 1/\sqrt{u_0}$, which translates into $s\lesssim$ 20 au for KB180029 and $s\lesssim$ 24 au for OB190960.
On the other hand, the fact that the source/blend are aligned to within the 70 mas PSF in the Keck images rules out wide binary companions of $s\gtrsim$ 100-200 au.
The resulting narrow range of projected separation allowed for the lens companion further reduces the plausibility of this scenario, which may nevertheless be further investigated via multi-epoch late-time AO observations.

The possibility of a source companion may be informed by the color and magnitude of the excess flux. Thanks to the absence of comparably bright field stars within 2'' of OB190960's location (Figure 3), the optical blend flux as originally reported by \cite{yee_ogle-2019-blg-0960_2021} must be associated with the event. The (V$-$I) color and magnitude of the blend are inconsistent with it being in the Bulge, thereby ruling out the source companion scenario. As for KB180029, no optical color information could be reliably determined given the presence of a bright field star 0.5'' east of the source (Figure 3). Therefore, the possibility that the Keck excess flux for KB180029 reflects a source companion requires future imaging epochs measuring the lens-source relative proper motion to be completely ruled out.

The lens physical properties may be further refined by resolving the lens and source stars in the near future.
Measurements of the magnitude and direction of the lens-source relative proper motion not only refine both the angular Einstein radius and microlensing parallax, but also serve to break the various modeling degeneracies (e.g., \citealt{terry_adaptive_2022}). Nevertheless, the generalized inner-outer degeneracy for KB180029 cannot be resolved by follow-up observations due to its exactness when the source trajectory crosses the star-planet axis at a 90-degree angle.

\subsection{Resolving the inner-outer degeneracy via ground concurrent observations with the Roman Space Telescope}
There exists an opportunity to occasionally break the inner-outer degeneracy from ground concurrent observations with the Roman Space Telescope.
The Roman Space Telescope will operate in a halo orbit around the Sun-Earth L2 Lagrange point, located approximately 0.01 au from Earth, which provides an ideal separation for measuring satellite parallax for Earth-mass planets \citep{gould_resolving_2003}.
As the magnification matching behavior under the inner/outer degeneracy is not global but local to the source trajectory \citep{zhang_ubiquitous_2022}, the inner and outer models derived from the Roman data would generally predict different light curves for a ground-based observer, thus providing an opportunity for resolving this degeneracy.

\begin{figure*}[t]
    \centering
    \includegraphics[width=\textwidth]{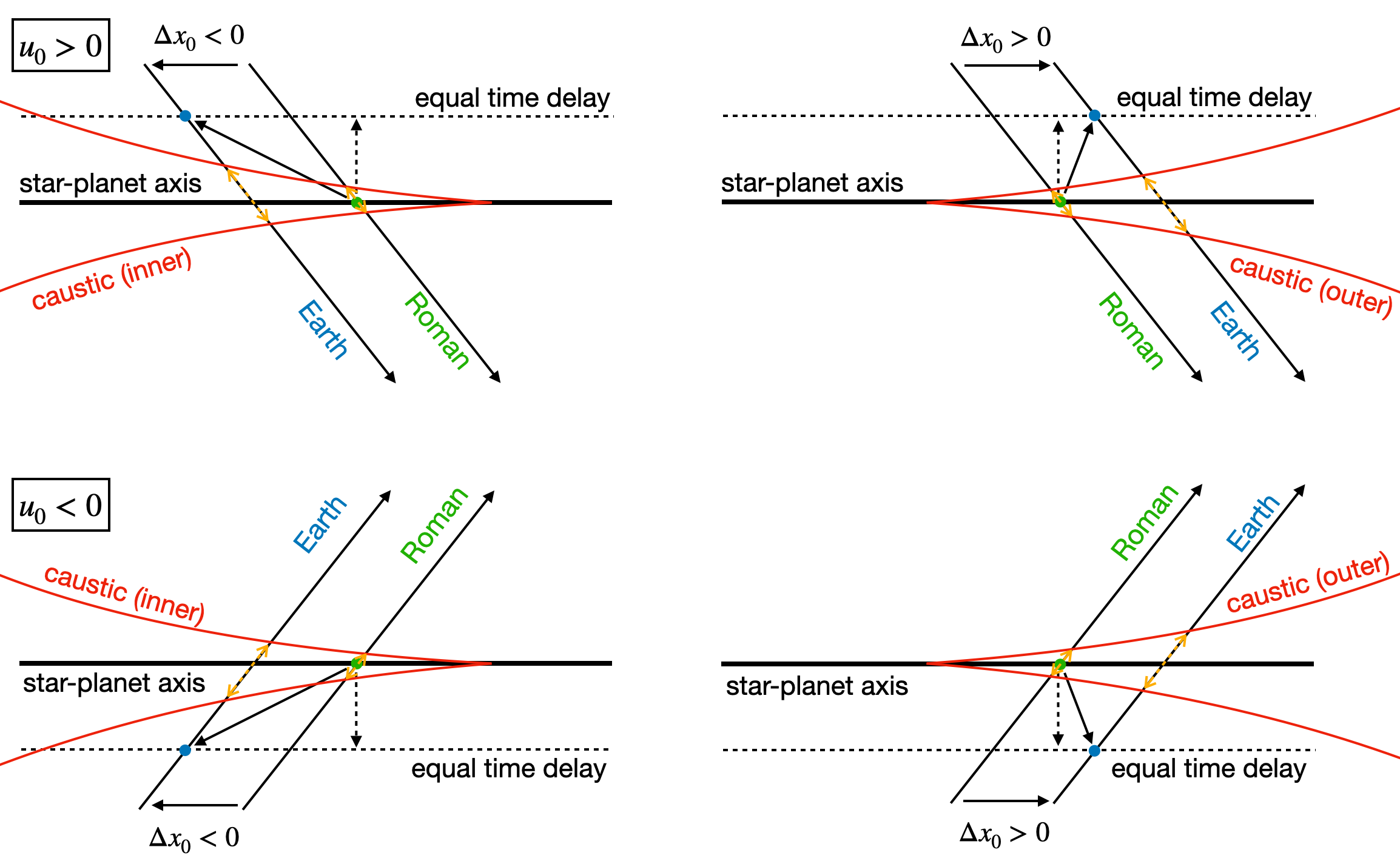}
    \caption{Illustration of a four-fold degeneracy for the Earth--Roman satellite parallax shown at the epoch when the source crosses the star-planet axis for Roman (green dot), assumed to be the time of planetary anomaly.
    The Earth-based source location is shown as the blue dot.
    Due to the inner-outer degeneracy, there are four possible configurations ($\pm \Delta x_0$; $\pm u_0$) with the identical Earth--Roman difference in the timing and duration of the planet anomaly.
    Within each scenario, the solid arrow indicates the source offset vector ($\Delta \boldsymbol u=\boldsymbol u_{\oplus}-\boldsymbol u_{\rm Roman}$), which lie along the on-sky projected Roman--Earth separation and whose magnitudes are proportional to the lens-source relative parallax.
    The vertical dashed arrow shows its projection perpendicular to the star-planet axis, which is informed by the anomaly time delay. Shifting the Earth-based source position along the horizontal dashed line does not change the time delay.}
    \label{fig:enter-label}
\end{figure*}

To illustrate this scenario, if KB180029 were observed by Roman, then the Earth-Roman baseline would shift the apparent source position relative to the lens by $0.01\cdot\pi_{\rm rel}\simeq0.004$ mas, at a lens distance of 2 kpc.
By comparison, the lens-source relative proper motion for KB180029 is 3.3 mas/yr, or 0.01 mas/day. 
If the Earth-Roman separation vector is projected perpendicular to the star-planet axis, the time of the planetary anomaly as seen from Earth and Roman would differ by approximately 10 hours, which could be measured by a medium cadence ground concurrent survey, provided that the magnified source is sufficiently bright.

In comparison, if the on-sky projected Earth-Roman separation vector is parallel to the star-planet axis, there would be no time delay in the planetary anomaly, but the source trajectory intercept on the star-planet axis, which is simply the impact parameter for KB180029, would differ by $|\Delta x_0|\simeq0.003$, causing a difference in the duration of the planetary anomaly as seen from the two vantage points.
This difference could be too small to be measured in practice, especially in the case of KB180029, where the caustic cross section changes little over the range $x_0 \pm \Delta x_0$ (Figure 1).
It is likely easier to measure when the source trajectory passes closer to the planetary caustic/cusp, where the caustic vertical cross section changes more rapidly.

As illustrated in Figure 5, the inner-outer degeneracy leads to a four-fold satellite parallax degeneracy for the general case where source trajectory does not cross the star-planet axis at a 90-degree angle. 
In this case, the inner-outer degeneracy is no longer exact, but let us assume that the Roman data alone does not break this degeneracy for the purpose of this analysis.
If differences in both the timing and duration of the planetary anomaly were measured, there would be a two-fold degeneracy in the direction of the source trajectory ($\pm u_0$), and a two-fold degeneracy in the direction of the source intercept offset ($\pm \Delta x_0$).
In the example shown in Figure 5, the planetary anomaly occurs later and for a longer duration as observed from Earth for all four configurations.

First, note that the anomaly time delay informs the satellite parallax component perpendicular to the star-planet axis, as moving the blue dot along the horizontal dashed line does not change the time delay.
Moreover, a change in the direction of the source trajectory ($u_0\rightarrow-u_0$) preserves the time delay, but flips the perpendicular parallax component to the opposite side of the star-planet axis. This results in a degeneracy in the sign of $u_0$ similar to the ecliptic degeneracy.

This 1D satellite parallax can generally be combined with the 1D annual parallax to derive the full 2D microlensing parallax, with the exception of vertical source trajectories  \citep{gould_resolving_2003} like KB180029.
The 1D annual parallax preferentially measures a projection parallel to the source trajectory \citep{gould_microlensing_1994}, which is also the direction perpendicular to the star-planet axis in this special case.

As for the measured difference in the anomaly duration, observe that the caustic structure under the inner and outer models are locally symmetric under a left-right flip about the vertical direction (Figure 1).
For the anomaly to be longer for a ground-based observer, the inner and outer models would predict intercept offsets ($|\Delta x_0|=|x_{\rm 0, \oplus}-x_{\rm Roman}|$) in opposite directions. This completes the four fold degeneracy.

Among the four configurations, there are two different magnitudes and four different directions of the satellite parallax relative to the source trajectory.
Therefore, since the direction of the Earth-Roman separation vector is known in advance, a late-time measurement of the direction of the lens-source relative proper motion may serve to resolve this degeneracy.

Alternatively, recall that the 1D annual parallax preferentially measures a projection parallel to the source trajectory. From Figure 5, the projection of the satellite parallax onto the source trajectory differs for the inner and outer ($\pm\Delta x_0$) models and is invariant to a sign flip in $u_0$.
Therefore, it may also be possible to break this degeneracy from a simultaneous measurement of the 1D annual microlensing parallax.
A more detailed consideration of the Roman satellite parallax is deferred to future work.

\begin{acknowledgements}
We thank Weicheng Zang for contributing to the Keck proposal, and Natalie LeBaron for assisting with observing. K.Z. is supported by the Eric and Wendy Schmidt AI in Science Postdoctoral Fellowship, a Schmidt Futures program. K.Z. and J.S.B. were partially supported by the Gordon and Betty Moore Foundation and a grant from the National Science Foundation (award \#2206744). S.K.T. is supported by NASA under award number 80GSFC24M0006. B.S.G. is supported by a Thomas Jefferson Chair for Discovery and Space Exploration Endowment and NASA grant 80NSSC24M0022. J.R.L. acknowledges support from the National Science Foundation under grant No. 1909641 and the Heising-Simons Foundation under grant No. 2022-3542. Some of the data presented herein were obtained at Keck Observatory, which is a private 501(c)3 non-profit organization operated as a scientific partnership among the California Institute of Technology, the University of California, and the National Aeronautics and Space Administration. The Observatory was made possible by the generous financial support of the W.\ M.\ Keck Foundation. The authors wish to recognize and acknowledge the very significant cultural role and reverence that the summit of Mauna Kea has always had within the Native Hawaiian community. We are most fortunate to have the opportunity to conduct observations from this mountain.
\end{acknowledgements}

\begin{contribution}
K. Zhang performed the majority of scientific analysis and wrote the manuscript.
S. K. Terry contributed to the PSF analysis.
S. K. Terry, J. S. Bloom, and J. R. Lu contributed to observing.
B. S. Gaudi discussed the satellite parallax.
All co-authors commented on the manuscript.
\end{contribution}

\end{CJK*}
\end{document}